\documentclass{PoS}

\title{Coulomb gauge ghost propagator and the Coulomb potential\thanks{Supported
by DFG under contract \textsl{Re-856/6-1,2}} }

\ShortTitle{Coulomb gauge ghost propagator and the Coulomb potential }

\author{\speaker{Markus Quandt}\\
        University of T\"ubingen
        \\
        E-mail: \email{quandt@tphys.physik.uni-tuebingen.de}}

\author{Giuseppe Burgio\\
        University of T\"ubingen\\
        E-mail: \email{burgio@tphys.physik.uni-tuebingen.de}}

\author{Songvudhi Chimchinda\\
        University of T\"ubingen and Suranaree  University of  Technology
        \\
        E-mail: \email{chimchinda@tphys.physik.uni-tuebingen.de}}

\author{Hugo Reinhardt\\
        University of T\"ubingen\\
        E-mail: \email{hugo.reinhardt@uni-tuebingen.de}}

\abstract{The ghost propagator and the Coulomb potential are evaluated
in Coulomb gauge on the lattice, using an improved gauge fixing scheme
which includes the residual symmetry. This setting has been shown
to be essential in order to explain the scaling violations in the
instantaneous gluon propagator. We find that both the ghost propagator
and the Coulomb potential are insensitive to the Gribov problem or the
details of the residual gauge fixing, even if the Coulomb potential is
evaluated from the $A_0$--propagator instead of the Coulomb kernel.
In particular, no signs of scaling violations could be found in either
quantity, at least to well below the numerical accuracy where these
violations were visible for the gluon propagator. The Coulomb potential
from the $A_0$-propagator is shown to be in qualitative agreement with the
(formally equivalent) expression evaluated from the Coulomb kernel.
}

\FullConference{8th Conference Quark Confinement and the Hadron Spectrum\\
                 September 1-6, 2008\\
                 Mainz. Germany}

\newcommand{\vek}[1]{\mathbf{#1}}

\begin{document}

\section{Introduction}
Yang--Mills theory in the Coulomb gauge has recently drawn a
renewed attention, both in the continuum \cite{szepaniak, claus_hugo, depple}
and on the lattice \cite{kurt, cucc, voigt, prop, mmp}.
This is mainly due to the fact that Gau{\ss}' law can be resolved explicitly in
this gauge, which allows for a neat Hamiltonian formulation with the transversal
part of the remaining vector potential $\vek{A}^\perp$ as the only physical
degree of freedom. Much of the intuition and techniques from ordinary quantum
mechanics can thus be carried over to the YM case. In particular, recent
variational approaches in the Schr{\"o}dinger picture, based on the notion of a
weakly interacting constitutent gluon and the Gribov--Zwanziger confinement
scenario \cite{zwanziger}, proved to be very successfull \cite{depple};
similar calculations are presently carried out in the renormalisation
flow approach.

All these continuum formulations, in one way or the other, give rise to
relations between low-order Green functions of the constituent gluon
$\vek{A}$ and the Faddeev--Popov ghosts. It is therefore important to obtain
non-perturbative information on such correlators from the lattice.
Careful studies of the equal--times gluon propagator, for instance, reveal
strong scaling violations and a UV behaviour at odds with simple dimensional
arguments \cite{kurt, prop, voigt}.
These surprising results reflect the renormalisation problems
for instantaneous correlators in the continuum. One possible explanation
of the lattice findings \cite{prop_prl} is based on the idea that the
residual gauge freedom left over by the Coulomb condition must be fixed
in such a way that it resembles the Hamiltonian formulation as closely
as possible.\footnote{For the first-order formalism in the continuum,
renormalisability has been proven algebraically \cite{renorm}.}
A careful study of the energy dependence of the gluon propagator then allows
to manipulate the data such that perfect scaling is observed even on finite
lattices.

For the confinement scenario layed out by Gribov and Zwanziger\cite{zwanziger},
the more important correlators are, of course, the ghost propagator and,
in particular, the Coulomb potential. Furthermore, the ghost form factor has
been shown to represent the inverse of the colour dielectric function of the
Yang--Mills vaccum \cite{hugo2}, and is therefore of direct
physical relevance. Initial studies of the ghost and Coulomb propagator
for the gauge group $G=SU(2)$ with simple Coulomb and no residual gauge
fixing \cite{kurt} found no scaling violations at low momenta, but had
inconclusive results about the Coulomb string tension in the deep infrared.
Moreover, these results were partially at odds with more careful $SU(3)$
studies using a residual gauge fixing different from ours \cite{voigt_su3},
which featured a peculiar saddle-like behaviour in the Coulomb potential at
low momenta.
In the present talk, I will report about recent $SU(2)$ calculations of
ghost form factors and the Coulomb potential, using exactly the same gauge
fixing techniques which proved essential for the resolution of the
scaling violations in the gluon propagator.

\section{Gauge Fixing}
Our gauge fixing procedure employs both simulated annealing and the
microcanonical flip procedure layed out in \cite{prop} as a preconditioning
with subsequent (over)relaxation to complete the gauge fixing within machine
precision. To reduce the Gribov noise and bring the lattice configs closer to
the fundamental modular region, we perform up to 40 restarts with random
gauge transformations as starting points, and take the copy with the best
minimum of the gauge fixing functional. While this procedure proved to be
important for the correct extraction of the gluon propagator in the
deep infrared \cite{prop}, the ghost correlators exhibit a much weeker dependence
on the quality of gauge fixing. This can be clearly seen in the left panel of
fig.~\ref{picghost}: The value of the ghost propagator at the lowest diagonal
momentum $\hat{\vek{p}} = (1,1,1)$ is only very slightly suppressed as the
number $n$ of Gribov restarts is increased, and the optimum is already reached for
$n$ as low as $n\approx 2..3$. All this is in constrast to the corresponding
findings for the gluon propagator, where a $20\%$ effect was seen that required
up to $n=40$ for saturation.

\begin{figure}
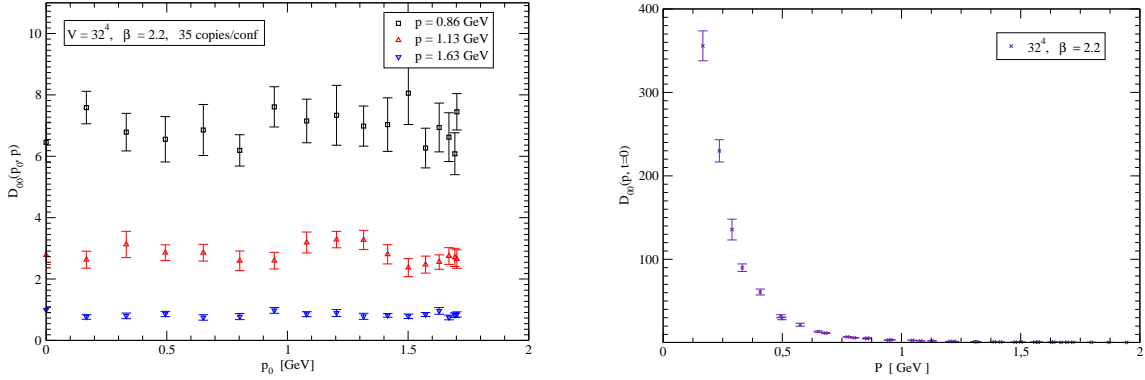

\vspace*{1mm}
\includegraphics[width=7cm]{D00_p0.eps}
\hfill
\includegraphics[width=7cm]{D00.eps}
\caption{Left panel: Energy dependence of the $A_0$--propagator
$D_{00}(\vek{p},p_0)$ after improved Coulomb and resiudal gauge fixing,
for various spatial momenta $|\vek{p}|$.
Right panel: The equal-times $A_0$--propagator $D_{00}(\vek{p}, t=0)$
as a function of the spatial momentum $|\vek{p}|$.}
\label{pica0}
\end{figure}

The second important ingredient is the residual gauge fixing. To make contact
with the Hamiltonian approach in Weyl gauge, we would like to put
the spatial average $u(t) = L^{-3}\sum_{\vek{x}} U_0(t,\vek{x})$ to unity.
However, periodic boundary conditions only allow us to make $u(t)$
time-independent, $u(t) \equiv \overline{U}_0 = \mbox{const}$.
In the infinite
volume limit (and in praxis also for $L \ge 32$), $\overline{U}_0$
approaches unity. Although this only enforces $\partial_0 U_0 = 0$ on the
spatial average, the $A_0$--propagator is, within statistical errors,
independent of energy (see left panel of fig.~\ref{pica0}). In the right
panel of fig.~\ref{pica0}, we thus plot only the instantaneous $A_0$--propagator
which is strongly enhanced in the infrared. This result will be related to
the Coulomb potential below.

\section{Results}

\begin{figure}
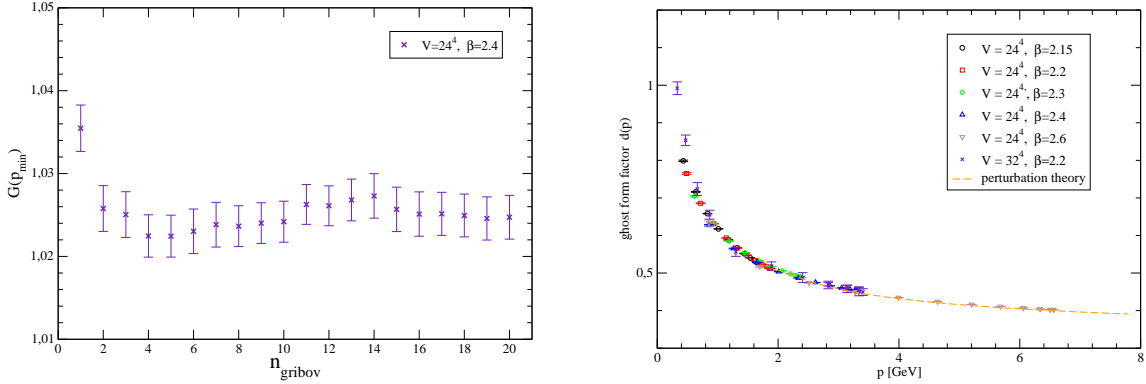

\vspace*{1mm}
\includegraphics[width=7cm]{ghost_gribov_noise.eps}
\hfill
\includegraphics[width=7cm]{ghost_form.eps}
\caption{Left panel: The ghost propagator at the lowest diagonal momentum
$\hat{\vek{p}} = (1,1,1)$ as a function of the number of Gribov copies
considered in the Coulomb gauge fixing. (Note the scale on the $y$--axis.)
Right panel: The ghost form factor $d(p)$ as a function of the spatial momentum
$|\vek{p}|$.}
\label{picghost}
\end{figure}

The right panel of figure \ref{picghost} shows our results for the ghost propagator
and its form factor,
\begin{equation}
G(p) = \langle\, \bar{c}(-\vek{p})\, c(\vek{p})\,\rangle =
L^{-3}\,\sum_{\vek{x}}\,e^{i \vek{p}\vek{x}}\,
\langle \,\mathsf{M}(\vek{x},\vek{0})^{-1}\,\rangle  \equiv
\frac{d(|\vek{p}|)}{ \vek{p}^2}\,
\end{equation}
where $\mathsf{M} \equiv (- \nabla \vek{D})$ is the Faddeev--Popov operator
and the ghost form factor $d(p)$ measures the deviation from the perturbative
result. The form factor is infrared enhanced, which agrees with the
horizon condition $d^{-1}(0) = 0$ necessary in the Zwanziger confinement
criterion \cite{zwanziger}. Our infrared exponent $\kappa \approx 0.22$
for the divergence $d(p) \sim 1/(p^2)^\kappa$ is slightly smaller than the
one obtained with naive gauge fixing \cite{kurt}, but agrees well with recent
improved studies in $SU(3)$ \cite{voigt_su3}.

Even more directly related to the confinement problem is the so-called
Coulomb potential $V_c$, i.e.~the response of the gluon vacuum to static colour
charges. Since the constituent gluon $\vek{A}$ and its wave functional are
gauge-dependent, $V_c$ is not directly the physical potential between
static quarks (as extracted from Wilson loops or Polyakov lines),
but an upper bound, $V_c(r) \ge \frac{4}{3} \,V(r)$.
This implies that there is no confinement without Coulomb confinement
\cite{zwanziger}, but a linear Coulomb potential may persist even in the
deconfined phase.

Formally, $V_c(r)$ can be computed in one of two equivalent ways,
\begin{equation}
V_c(|\vek{x}-\vek{y}|) = \langle \,A_0(t,\vek{x})\,A_0(t,\vek{y})\,\rangle
= g^2\,\langle \,\left(\mathsf{M}^{-1}\cdot \Delta\cdot \mathsf{M}^{-1}
\right)_{\vek{x},\vek{y}}\,\rangle\,.
\label{2}
\end{equation}
The formal equivalence of these two expressions can be shown in the
first order formalism upon explicitly resolving Gau{\ss}' law
\cite{renorm, peterhugo}. This leaves
possible renormalisation issues aside and the lessons learned from the
scaling violations in the gluon propagator indicate that some caution is
required when connecting bare instantaneous correlators. Of course, the
$A_0$--propagator is numerically much simpler than the complicated
Coulomb kernel involving two inversions of the FP operator.

The strong Ward identities in Coulomb gauge \cite{renorm} imply that the
special combination in momentum space
\begin{equation}
\vek{p}^2\, V_c(\vek{p}) \sim g^2(p)
\end{equation}
is a renormalisation group invariant which can be taken as a definition of the
running coupling constant. Simulations with different $\beta$ should thus
fall on top of each other without further multiplicative renormalisation.
We have tested this conjecture for numerous values of $\beta$ on relatively
small $16^4$ lattices. (On $32^4$ lattices, we have only been able to complete
the analysis of the complicated Coulomb kernel for a single value of $\beta$).
The $\beta$--invariance was much better for the $A_0$--correlator, while
$V_c$ constructed from the Coulomb kernel still showed noticable
scaling violations. At present, it is not known whether these deviations
are pure numerical or finite volume effects, or if they have any more significant
meaning. (Similar observations were made in ref.~\cite{voigt_su3}).
Simulations with improved statistics on larger lattices have to be conducted
to resolve this issue.

Finally, the most direct approach to the confinement issue is
given by the expression
\begin{equation}
\vek{p}^4  \,V_c(\vek{p})\,.
\label{4}
\end{equation}
From the Fourier transformation of a linear potential, $V_c(r) = \sigma_c\,r$,
it is readily seen that
\[
\vek{p}^4  \,V_c(\vek{p}) \to 8 \pi \sigma_c\,,\qquad\quad |\vek {p}| \to 0\,.
\]
The \emph{Coulomb string tension} $\sigma_c$ is an upper bound for the real
string tension $\sigma$ extracted from Wilson loops. Previous and current
lattice studies are inconclusive as to whether $\sigma_c = \sigma$,
since the approach to $|\vek{p}|\to 0$ is not as uniform as expected:
Early simulations without improved/residual gauge fixing saw a slight but
noticeable rise in the quantity (\ref{4}) below
$|\vek{p}|\approx 1\,\mathrm{GeV}$, which seemed compatible with
$\sigma_c / \sigma$ anywhere in the range $1\ldots3$. More recent computation
for the gauge group $G=SU(3)$ prefer a value $\sigma_c / \sigma \approx 1.6$,
but the extrapolation to zero momentum is again uncertain due to a peculiar
"bump" in the quantity (\ref{4}) at momenta between $0.1\ldots 1\,\mathrm{GeV}$.

\begin{figure}
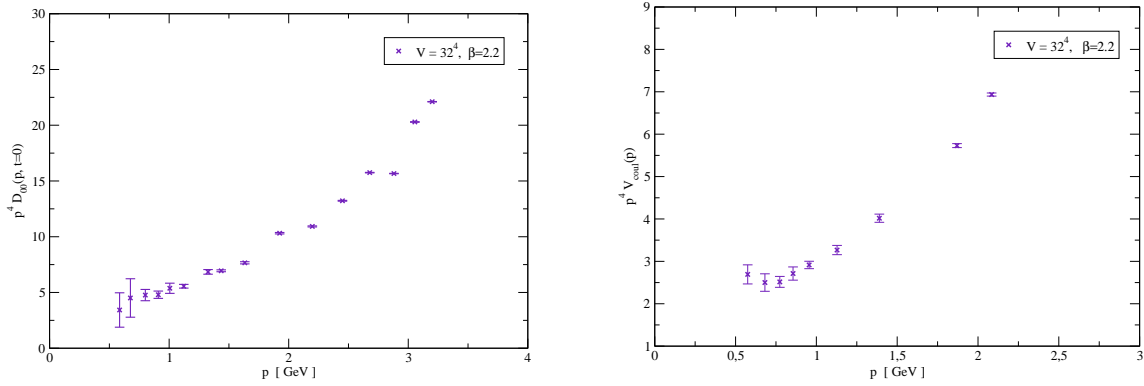

\vspace*{1mm}
\includegraphics[width=7cm]{p4_D00.eps}
\hfill
\includegraphics[width=7cm]{p4_V_32.eps}
\caption{Left panel: The combination $\vek{p}^4\,V_c(|\vek{p}|)$
with the Coulomb potential $V_c$ extracted from the $A_0$--propagator
$D_{00}(\vek{p},t=0)$.
Right panel: The same quantity, with $V_c$ extracted from the Coulomb
kernel.}
\label{piccoulomb}
\end{figure}

Our result on a $V=32^4$ lattice in figure \ref{piccoulomb}
using all improved gauge fixing techniques give reliable results
(for cylinder cut momenta) only down to
$|\vek{p}| \simeq 0.5\,\mathrm{GeV}$. In this range, the results for
(\ref{4}) are compatible with $V_c$ computed either from the $A_0$--propagator
or from the Coulomb kernel. The latter result show a more pronounced plateau
at the smallest momenta, which is reminiscent of the slight rise observed
in \cite{kurt}. However, the numerical data can equally well be fitted with
a constant. ($V_c$ from the $A_0$--propagator is compatible with the Coulomb
kernel results within statistical errors).
For both definitions of $V_c$, we do not see the "bump" reported for
$SU(3)$ in ref.~\cite{voigt_su3}. While the approach to a constant seems
promising, better statistics and larger lattices are required for a reliable
extrapolation of $\sigma_c/\sigma$.

\section{Conclusion}
The computation of ghost correlators and the Coulomb potential in $G=SU(2)$
show qualitative agreement with continuum calculations in the variational
approach \cite{claus_hugo, depple}. The scaling violations observed previously
for the equal-times gluon propagator $D(\vek{p})$ have no counter part in the
ghost correlators studied here. In particular, the dependence on the Gribov
noise and the details of the improved gauge fixing are negligable.
Likewise, the residual gauge
fixing, which is essential for the resolution of the scaling violations  in
$D(\vek{p})$, seems to have little or no influence on the ghost propagator
or Coulomb potential, even when the latter is extracted from the the
$A_0$--propagator.

Our residual gauge fixing removes the energy dependence on $A_0$ not only in
the spatial average, but effectively for arbitrary $A_0$--correlators. There
is thus no issue with renormalisation and the results for the instantaneous
$A_0$--correlators resemble the ones with unfixed residual symmetry.
(Similar observations are made for the ghost propagator and the Coulomb potential
as extracted from the Coulomb kernel.) It is therefore not surprising that
our findings agree with other calculations, even if these fixed the Coulomb
gauge naively, or left the residual symmetry unfixed.

The statistics in the deep infrared are not sufficient to make reliable
quantitative extrapolations for the Coulomb string tension $\sigma_c$,
or the Coulomb form factor $f(p)$ whose infrared behaviour is an important
ingredient in the variational approaches \cite{claus_hugo, depple}.
We intend to improve on this and accumulate data for $32^4$ lattices with
various $\beta$, and $A_0$--correlators on even larger lattices.
These results will be published in a forthcoming paper.

\end{document}